# Reduced-Graphene-Oxide with Dispersed Au-Ir Nanoparticles as Active Support for Pt at low Loadding for Electrocatalytic Oxygen Electroreduction

*Sylwia Zoladek*[a]*, *Magdalena Blicharska*[a], *Anna Jablonska*[a,b], *Iwona A. Rutkowska*[a],
*Cezary Guminski*[a], *Krzysztof Miecznikowski*[a], *Maciej Krzywiecki*[c], *Jerzy Zak*[d], *Enrico Negro*[e],
*Vito Di Noto*[e], *B. Palys*[a,b], *Pawel J. Kulesza*[a]*

[a]Faculty of Chemistry, University of Warsaw, Pasteura 1, PL-02-093 Warsaw, Poland

[b]Biological and Chemical Research Centre, University of Warsaw, Żwirki i Wigury 101, 02-089 Warsaw, Poland

[c]Institute of Physics CSE, Silesian University of Technology Konarskiego 22B, 44-100 Gliwice, Poland

[d]Department of Physical Chemistry and Technology of Polymers, Silesian University of Technology, Strzody 9, 44-100 Gliwice, Poland

[e]Department of Industrial Engineering, Università degli Studi di Padova in Department of Chemical Sciences, Via Marzolo 1, 35131 Padova (PD), Italy

Corresponding authors:
E-mail: szoladek@chem.uw.edu.pl; pkulesza@chem.uw.edu.pl

**Keywords:**

Chemically reduced graphene oxide, gold, iridium, platinum nanoparticles; Electrocatalysis; Oxygen reduction; acidic medium

**Abstract**

We report here on a novel and facile techniques for the synthesis nanocomposite based on stable bimetallic catalyst containing iridium and gold nanoparticles electrodeposited on chemically reduced graphene oxide (rGO) sheet admixed with platinum nanoparticles as an efficient electrocatalyst to facilitate the oxygen reduction reaction (ORR) in acidic medium. Raman spectroscopy, infrared spectroscopy (IR), scanning electron microscope (SEM), transmission electron microscope (TEM) methods were employed to characterize the rGO, Au-NPS-rGO, Ir-NPS-rGO, Au-Ir-rGO nanocomposites formed on the electrode surface. Here we present a comparative study into the relative effects of each hybrid nanocomposites: Au-NPS-rGO, Ir-NPS-rGO, Au-Ir-rGO nanocomposites on the performance of platinum nanoparticles towards the oxygen reduction reaction (ORR) in acidic media. For the purposes of comparing the supports, a simple platinum black catalyst is used and the performance is evaluated via direct measurement of peroxide by rotating ring-disk electrode (RRDE) to determine the number of electrons (n) transferred in the ORR. It was found that platinum nanoparticles dispersed within Au-Ir-rGO support follows a quasi 4-electron mechanism (3.99-4 electrons involved) due to that the ORR takes place mainly on the active Pt particles, and produced hydrogen peroxide is reduced at Au-Ir-rGO support.

# 1. Introduction

Recently, renewable energy devices operated by electrochemical principles attracted concentrated attention especially because of not only its simple design but also high performance [1]. Proton exchange membrane fuel cell (PEMFC), systems are believed to provide solutions to the energy sustainability and environmental pollution problems. It was demonstrated that among representative systems the PEMFCs can produce a large-scale electricity and power portable electronic devices and vehicles [2-5].

So far, the most efficient catalysts known for ORR are based on platinum (Pt), nevertheless, due to high cost of the noble metal, there has been growing interest in minimizing its content in catalytic layer while iridium (Ir) and ruthenium (Ru) oxides are considered as the most active catalysts for OER and hydrogen peroxide electroreduction. However, they only have moderate activities for the reverse reaction [6-9]. Substantial research efforts have been directed to decreasing content of Pt and Ir catalysts with high catalytic activity and good durability [10, 11]. One of possible approaches to achieve this goal is to utilize robust conducting (electronically and ionically) matrices (e.g., certain metal oxides) which are capable of reducing the rate of particles agglomeration and degradation, would interact specifically with dispersed noble metal centers to affect their structural, electronic, chemisorptive, and interfacial properties and are also highly reactive towards reduction of hydrogen peroxide (i.e., the undesirable oxygen reduction intermediate) [11]. Moreover bimetallic catalysts are of great importance to oxygen electroreduction process [12-15]. Recently, gold-based bimetallic catalysts have attracted much attention owing to their unique catalytic properties. The addition of a second metal causes changes of the electronic properties of gold particles, and change their local atom distribution by forming core-shell structures between gold and the added metal [16,17]. Iridium atoms are known to donate electrons more easily than gold atoms do; therefore, modification of gold with iridium atoms courses interactions that change the ability of gold to donate electrons [18]. A very limited miscibility in bulk, hinders the possible formation of alloys with randomly dispersed Ir and Au atoms in the same crystalline structure[19]. According to recent study of supported Au-Ir catalysts gold atoms could cover the surface of iridium particles as metallic gold has a lower surface free energy (1410 erg cm$^{-2}$) than metallic iridium (3000 erg cm$^{-2}$) does gold atoms are more likely to dominate the bimetallic surface [18]. Although the Au and Ir in the AuIr/C do not form alloy, it is clear that the introduction of Ir decreases the average Au particle size to 4.2 nm compared to that in the Au/C (10.1 nm) [19]. Moreover the combination of Au and Ir, protects Au against being oxidized due to the lower electronegativity of Ir. Thus combining the advantages of Au and Ir in catalyzing ORR and OER, the AuIr/C catalyst displays an enhanced catalytic activity to the ORR and a comparable OER activity. Gomez-Cortes and co-workers reported that Au–Ir catalysts prepared by sequential deposition (first depositing Ir and then Au) showed higher catalytic activity for the oxidation of CO than those containing only gold and were more stable with the active sites being small particles containing gold and metallic iridium [20].

In the present study, we consider three-dimensional (3D) crumpled reduced graphene oxide decorated with Au-Ir or Au or Ir nanoparticles as active supports for platinum nanoparticles active in electrocatalytic oxygen electroreduction process. The multifunctional nanocomposites: Au-NPS-rGO, Ir-NPS-rGO, Au-Ir-rGO were synthesized by electrochemical deposition method. Physical characterizations as: Raman spectroscopy, infrared spectroscopy (IR), transmission electron microscope (TEM) and scanning electron microscopy (SEM) with (EDX) analysis was conducted to examine their properties. The electro catalytic properties of the resulting materials toward oxygen electroreduction were investigated using rotating ring-disk electrode (RRDE).

**2. Experimental**

Chemicals were commercial materials of the highest available grade, and they were used as received. $H_2SO_4$, ethanol and $K_3[Fe(CN)_6]$ were obtained from POCh (Poland). The hydrogen tetrachloroaurate (III) trihydrate HAuCl$_4$·3H$_2$O (>99.9%); sodium borohydride (powder, 98%), NaBH$_4$;, 5 wt% Nafion solution were purchased from Sigma-Aldrich and were used without any further purification. The IrCl$_3$·xH2O was from Alfa Aesar (Germany). Nitrogen and oxygen gases (purity 99.999%) were from Air Products (Poland). Graphene oxide (GO) was synthesized from graphite powder (Sigma-Aldrich; 1-2 μm) using the modified Hummers-Offeman method [21, 22].

The spectral Raman profiles were measured with using a DXR Raman spectrometer (Thermo Scientific). The instrument was operated using a 780-nm excitation line. For typical experiments, the spectral resolution was equal to 1 cm$^{-1}$. The 50 × /NA0.75 objective was used. In the configuration used, the signal was collected from the 1- μm$^3$ spot. The laser beam was focused on top of the layer to minimize the contribution of the GC support to the spectra.

The infrared spectra were taken with Thermo Scientific, model Nicolet iS50 FT-IR. Morphology of samples was assessed using Libra Transmission Electron Microscopy120 EFTEM (Carl Zeiss) operating at 120 kV.

Electrochemical measurements were carried out on CH Instruments (Austin, TX, USA) Models: 600B and 750A workstations. The electrochemical cell was assembled with a conventional three-electrode system: a rotating ring disk working electrode (RRDE), saturated calomel electrode (SCE) reference electrode (exhibiting potential of ca. 240 mV relative to the reversible hydrogen electrode (RHE)) and carbon wire counter electrode. The rotating ring disk electrode (RRDE) working assembly was from Pine Instruments; it included a glassy carbon (GC) disk and a Pt ring. The radius of the GC disk electrode was 2.5 mm; and the inner and outer radii of the platinum ring electrode were 3.25 and 3.75 mm, respectively. Before experiments, working electrode was polished with aqueous alumina slurries (grain size, 5-0.05 mm) on a Buehler polishing cloth. Later, the glassy carbon disk electrode was subjected to potential cycling in 0.5 mol dm$^{-3}$ H$_2$SO$_4$ for 20 min in the potential range from 0 to 0.9 V. The Pt ring electrode was also subjected to potential cycling in the same solution but in the range of the potential form 0 V to 1.2 V. The collection efficiency (N) of the RRDE assembly, was calibrated by $K_3Fe(CN)_6$ redox reaction determined according to procedure described before [23]. During the RRDE experiments in oxygen saturated solutions, the potential of the ring electrode was kept at 1.2 V vs. RHE. The activity of the prepared catalytic films was evaluated by performing voltammetric potential cycles in the range from 0.1 to 1.1 V vs. RHE at 50 mVs$^{-1}$ until stable voltammetric responses were observed at a scan rate of 10 mVs$^{-1}$ under nitrogen atmosphere. Before the electrochemical measurement the electrolyte was purged and saturated with $N_2$ or $O_2$ gas. A constant nitrogen (or oxygen) flow over the solution was maintained during all measurements. All RRDE polarization curves were recorded at the scan rate of 10 mV s$^{-1}$, typically with a rotation rate of 1600 rpm. The ring potential was maintained at 1.21 V vs RHE to oxidize any hydrogen peroxide produced. Experiments were performed at room temperature (22 ± 2ºC).

Fabrication of gold nanoparticles or iridium nanorods or Au-Ir particles deposited onto electrochemically reduced graphene oxide was performed according to electrochemical reduction of precursor solution dropped onto electrochemically reduced graphene oxide. In order to obtain the electrochemically reduced graphene oxide the 0.1 g of graphene oxide was suspended in the ethanol and, subsequently, subjected to 24 h stirring to obtain homogenous mixture with concentration of GO equal to 10 mg/ml. As a rule, appropriate amounts of the resulting inks was dropped onto surfaces of glassy carbon electrodes to obtain loadings of graphene oxide equal to 80 μg cm$^{-2}$, later, subjected to drying in air at room temperature for 20 min. Then GO reduced by cycling the electrode potential in the range from−0.16 V to −0.96 V at 5 mV/s in 0.5 mol dm$^{-3}$ $H_2SO_4$. Iridium or gold or Ir-Au nanoparticles were obtained by electrochemical reduction of appropriate amounts of the resulting precursor water solution dropped on electrochemically reduced graphene oxide. Each precursor was dropped onto reduced graphene oxide and subjected to drying in air at room temperature to obtain desired loadings of metallic nanoparticles equal to 5 μg cm$^{-2}$ Au NPs. Then layers were subjected to electrochemical reduction at the potential of -0.9 V vs. RHE from the solution of 0.5 mol dm$^{-3}$ H$_2$SO$_4$ for 0.5 s.

A typical catalytic ink containing Pt nanoparticles was prepared according to the procedure as follows. The appropriate volume (177.5 μdm3) of aqueous suspension of platinum nanoparticles mentioned above was dispersed in water containing 816.5 μdm3 of ethanol and mixed on the magnetic stirrer for 6 h. Then alcoholic solution of Nafion was added to the resulting suspension of platinum. The ink was magnetically mixed for the next 22 h. As a rule, an appropriate amount of the resulting ink was dropped on the: rGO or Au-NPS-rGO or Ir-NPS-rGO or Au-Ir-rGO to obtain platinum loading equal to 30 μg cm$^{-2}$.

## 3. Results and discussion

The Scanning Electron Microscopy image of the electrochemically reduced Graphene Oxide nanosheets obtained by the electrochemical reduction of graphene oxide is shown in Fig. 1. It was found that reduced GO nanosheets consists of randomly aggregated and crumpled thin sheets which a graphene surface that was crumpled in random shapes. In general, graphene nanosheets were crumpled to a curly and wavy shape and on its surface there were observed wrinkles and folds.

The characteristic FTIR spectrum of GO nanosheets is presented in Fig. 2a and compered to electrochemically reduced graphene oxide (Gr) (Fig. 2b). It is seen with oxygen containing groups in which the main absorption band at 3356 cm$^{-1}$ and 3180 cm$^{-1}$ is assigned to the O-H group stretching vibrations. The absorption peaks at 1715 cm$^{-1}$ and 1708 cm$^{-1}$ can be assigned to C=O stretching of carboxylic acid. The absorption peaks at about 1626 cm$^{-1}$ and 1581 cm$^{-1}$ are assigned to the OH bending. absorption band at 1378 cm$^{-1}$ is assigned to the in plane C-OH bending and 1224 cm$^{-1}$ and 874 cm$^{-1}$ should be assigned to the C-O-C bending. The absorption peaks at 1056 cm$^{-1}$ and 1054 cm$^-$can be attributed to the C-O (alkoxy) vibration modes [24]. It should After reduction, the amount and intensity of bands corresponding to the oxygen groups (especially epoxies) decreases (Fig 2a, 2b).

Fig. 3a shows representative Raman spectra of graphene oxide (GO) compared to electrochemically reduced graphene oxide (Gr) (Fig. 3b ). In particular, the Raman spectra show two large peaks: one located near 1330 cm$^{-1}$ which is attributed to the D band originating from the amorphous structures of carbon, and the second one close 1600 cm$^{-1}$, which is correlated with the G band and reflects the graphitic structures of carbon. It should be noted that after electroreduction process the bandwidth narrows. $I_D/I_G$ ratio of Gr and higher when compared to GO, respectively what indicates indicates that the surface concentration of interfacial defects on the Gr sheets is significantly higher when compared to Gr [25].

Morphology studies of chemically reduced graphene-supported gold nanoparticles catalysts carried out with transmission electron microscopy (Figure 1b) have revealed the presence of gold nanoparticles sized from 15 to 65 nm within the representative graphene sheet (Figure 1b). The result indicates that gold nucleation occurs mostly at dispersed defect sites of carbon support, including surface polar groups. The partly reduced interfacial layers of $PMo_{12}O_{40}^{3-}$ sites moderate reducing capability of carbon support providing excessive nucleation sites, what results in generation of bigger gold nanoparticles. Transmission electron microscopic images of reduced-graphene-oxide supported (A) gold nanoparticles (B) iridium nanotubes and gold- iridium nanostructures Au-Ir. Histograms A', B', and C' display distribution of sizes of the respective metallic nanostructures.

Transmission electron microscopic images of: Au-NPS-rGO, Ir-NPS-rGO, Au-Ir-rGO were shown on Fig. 4. TEM of reduced-graphene-oxide supported gold nanoparticles (Fig.4A), shows separated spherical NPs with a relatively narrow size distribution between 1 and 11 nm.

Representative TEM micrographs of the Au-NPS-rGO was presented in Fig. 4A. The obtained gold nanoparticles are uniform sized spheres with relatively rounded edges with an average size about 7nm. A high degree of uniformity in diameter and length is observed for the iridium nanoroads deposited onto reduced graphene oxide (Fig.4B) (Ir-NPS-rGO). The rods are observed to be close to 26 nm and 90 nm dimensions. The rod edges are observed as symmetrically rounded. From Fig. 4C we see a uniform coating of spherical Au-Ir nanoparticles on the Gr sheet. An inset in Fig. 4C' shows the histogram of the diameter of 8 nm.

EDX analysis (Fig. 5) was also used to confirm the formation of the Ir-NPS-rGO nanocomposite. The results showed the presence of carbon, gold and irydium in Au-Ir-rGO. The presence of additional elements in the analyzed structures was also confirmed by application of other methods of measurements, for example XPS analysis. EDX maps are shown in Figure 5b; carbon, oxygen, gold and iridium are detected on the surface of electrochemically reduced graphene oxide covered with Au-Ir nanoparticles. The amount of elements detected by using various methods is summarized in Table 5b. It should be noted that the ratio of Au:Ir is close to 1:1.

To comment on the influence of gold nanoparticles on the electrochemical characteristics of chemically-reduced graphene oxide the voltammetric experiments (Fig. 6A) have been performed. The dashed lines in Fig. 6A stand for the responses of bare rGO. Incorporation of gold nanoparticles (Fig. 6A), introduction of rGO leads to the small increase of background currents originating from the double-layer-type charging/discharging effects occurring at potential 0.6V attributed to the redox of carbons (pseudocapacitance). In the respective potential range, the responses characteristic of redox transitions leading to the formation and reduction of gold oxides, will not be observed. Fig. 6B represents the cyclic voltammetric responses of the nanostructured Au-Ir particles electrodeposited onto electrochemically reduced graphene oxide, compared to the dash line standing for iridium nanostructures deposited onto rGO registered in nitrogen saturated 0.5 moldm$^3$ H$_2$SO$_4$.

Due to low content of iridium nanoparticles both in case of Ir-NPS-rGO and Au-Ir-rGO two sets of well-developed peaks at potentials below 0.35 V, correspond to weakly (-0.1 V) and strongly (-0.25V) adsorbed hydrogen atoms at the Ir surface are particularly not observe (Fig. 6B.). It should be noticed that a gradual rise in oxidation currents (responsible for the Ir oxide/hydroxide formation is observed at potentials starting soon after the hydrogen region, especially for Au-Ir-rGO (Fig. 6B, Curve a) [9].

To get further into activating interactions between catalytic components the specific reactivity towards the H$_2$O$_2$ intermediate was measured using rotating disk voltammetric responses for the reduction of hydrogen peroxide (0,1 mmoldm$^{-3}$) recorded at glassy carbon electrode modified with: (A) Au-Ir-rGO, (B) Ir-NPS-rGO, (C) Au-NPS-rGO nanocomposites in deaerated 0.5 moldm$^{-3}$ H$_2$SO$_4$ (Fig.7). In this context, it can be concluded under RDE voltammetric conditions the Au-Ir-rGO system is characterized by the highest hydrogen reduction currents and the most positive potential shift , in comparison to the performance of single component systems (Ir-NPS-rGO, Au-NPS-rGO). It is noteworthy that effective reductive decomposition of H$_2$O$_2$ proceeding on Au-Ir-rGO is very promising for further O$_2$ electroreduction process.

Figure 8 shows the representative Scanning Electron Microscopy picture of platinum nanoparticles dispersed within Reduced-Graphene-Oxide decorated with Au-Ir Nanoparticles (RGO/AuNPs-IrNPs/PtNPs). Microscopic results implied high porosity as well as the granular morphology of the resulting catalytic layers which seems to be very promising for further electrocatalytic performances.

To investigate the electrocatalytic activities of as prepared catalysts: Pt-Au-NPS-rGO, Pt-Ir-NPS-rGO, Pt-Au-Ir-NPs-rGO, the hydrodynamic voltammetry (RRDE) experiments were carried out for the reduction of oxygen for $O_2$- saturated 0.5 mol dm$^{-3}$ $H_2SO_4$ at 1600 rpm rotation rate and 10 mVs$^{-1}$ (Fig 9). Normalized (background subtracted) rotating disk voltammogram for oxygen reduction at platinum nanoparticles dispersed within nanocomposites of Au-Ir-rGO, produced the higher current densities at the disk electrode when compared to the Pt-Au-rGO, Pt-Ir-NPS-rGO, and Pt-rGO. It should be emphasized that ring currents corresponding to the oxidation of $H_2O_2$ intermediate are the lowest for the Pt-Au-Ir-NPs-rGO (Figure 10). What is more the system utilizing Pt-Au-Ir-NPs-rGO is capable at driving the oxygen reduction reaction at much more positive potentials with relatively higher electrocatalytic currents at the disk electrode in comparison to the activity of Pt-Au-rGO, Pt-Ir-NPS-rGO, and Pt-rGO (Figure 10).

Fig. 11 shows the percentage amount of $H_2O_2$ (%$H_2O_2$) formed during electroreduction of oxygen under the conditions of RRDE voltammetric experiment for: Pt-Au-Ir-NPs-rGO, Pt-Au-rGO, Pt-Ir-NPS-rGO, and Pt-rGO [26, 27]:

$$X_{H_2O_2} = \frac{200 I_r/N}{I_d + I_r/N}$$

where $I_r$ and $I_d$ are the ring and disk currents, respectively of the RRDE electrode and N is the collection efficiency. The results clearly show that the production of $H_2O_2$ is the lowest for system utilizing Pt-Au-Ir-NPs-rGO (Figure 11, solid line).

The overall number of electrons exchanged per $O_2$ molecule (n) was calculated as a function of the potential under the conditions of RRDE voltammetric experiment for the systems utilizing: Pt-Au-rGO, Pt-Ir-NPS-rGO, and Pt-rGO (Figure 12, solid line), according to the equation given below [28]:

$$n = \frac{4 I_d}{I_d + I_r/N}$$

It was found that platinum nanoparticles dispersed within Au-Ir-rGO support follows a quasi 4-electron mechanism (3.99-4 electrons involved) due to that the ORR takes place mainly on the active Pt particles, and produced hydrogen peroxide is reduced at Au-Ir-rGO support.

## 4. Conclusions

By incorporation of platinum black into different hybrid films compoused of Au-NPS-rGO, Ir-NPS-rGO, Au-Ir-rGO and rGO the electrocatalytic activity of Pt nanoparticles towards electroreduction of oxygen has been enhanced. Remarkable increases of electrocatalytic currents measured under rotating disk-disk experiments for Au-Ir-rGO/Pt have been observed. The most likely explanation takes into account improvement of overall conductivity (due to the presence of graphene support) at the electrocatalytic interface, as well as and possibility of specific Au-Ir electronic interactions and existence of active hydroxyl groups on rGO surfaces in the vicinity of catalytic Pt sites. The hybrid supports: Au-NPS-rGO, Ir-NPS-rGO, Au-Ir-rGO are highly active towards reduction of hydrogen peroxide. Although Ir and Au are known to be thermodynamically immiscible in this work we show that the thermodynamic of the Ir and Au binary system can be changed by the nanoscale effects and the electrochemically synthesized Ir-Au NPs compound can have a solid-sold structure. In the proposed combination Ir protects Au against being oxidized due to the lower electronegativity of Ir. Combining the advantages of Au and Ir in catalyzing ORR the AuIr/Graphene catalyst displays an enhanced catalytic activity to the hydrogen reduction reaction when compared to bare Au/Graphene and Ir/Graphene catalysts. Under voltammetric conditions the hybrid material based on Ir/Au/Graphene/Pt displays an impressive eletrocatalytic performance toward ORR and shows the highest number of exchanged electrons (n). The results from the rotating ring disk electrode methods showed that Ir/Au/Graphene/Pt displayed a satisfied stability, suggesting that the Ir/Au/Graphene/Pt is a potential acid resistant bifunctional catalyst for the oxygen.

**Acknowledgments**

This work was supported by the European Commission through the Graphene Flagship - Core 1 Project (GA-696656). The Polish side appreciates support from National Science Center (Poland) under Maestro Project 2012/04/A/ST4/00287.

**Figure captions**

**Fig. 1** Scanning Electron Microscopy picture of electrochemically reduced Graphene Oxide.

**Fig. 2** Ir spectra of the representative samples of: (a) graphene oxide (GO) and (b) electrochemically reduced graphene oxide (Gr).

**Fig. 3** Representative Raman spectra of: (a) graphene oxide (GO) and (b) electrochemically reduced graphene oxide (Gr).

**Fig. 4** Transmission electron microscopic images of reduced-graphene-oxide supported (A) gold nanoparticles (B) iridium nanotubes and gold- iridium nanostructures Au-Ir. Histograms A', B', and C' display distribution of sizes of the respective metallic nanostructures.

**Fig. 5** Elemental energy-dispersive X-ray spectroscopy for reduced graphene oxide-supported- Au-Ir nanoparticles including: (EDX) analysis (a), spectrum map (b) and mapping images of Au (c) and Ir (d).

**Fig. 6** Cyclic voltammetric responses of the following catalytic systems: (A) gold nanoparticles deposited onto electrochemically reduced graphene oxide; dash line stands for gold free electrochemically reduced graphene oxide, (B) nanostructured Au-Ir particles electrodeposited onto electrochemically reduced graphene oxide, dash line stands for iridium nanostructures deposited onto electrochemically reduced graphene oxide. Electrolyte: nitrogen saturated 0.5 moldm$^3$ H$_2$SO$_4$. Scan rate: 10 mVs$^{-1}$.

**Fig. 7** Normalized (background-subtracted) rotating disk voltammetric responses for the reduction of hydrogen peroxide (0,1 mmoldm$^{-3}$) recorded at glassy carbon electrode modified

with: (A) Au-Ir-rGO, (B) Ir-NPS-rGO, (C) Au-NPS-rGO nanocomposites; Electrolyte: deaerated 0.5 moldm$^{-3}$ H$_2$SO$_4$. Scan rate: 10 mVs$^{-1}$. Rotation rate: 1600 rpm.

**Fig. 8**. Scanning Electron Microscopy picture of platinum nanoparticles dispersed within Reduced-Graphene-Oxide decorated with Au-Ir Nanoparticles (RGO/AuNPs-IrNPs/PtNPs).

**Fig. 9** Normalized (background subtracted) rotating disk voltammograms for oxygen reduction at platinum nanoparticles dispersed within nanocomposites: (A) Au-Ir-rGO, (B) Ir-NPS-rGO, (C) Au-NPS-, (D) rGO. Electrolyte: oxygen-saturated 0.5 moldm$^{-3}$ H$_2$SO$_4$. Scan rate: 10 mVs$^{-1}$. Rotation rate: 1600 rpm.

**Fig. 10** Normalized (background subtracted) rotating ring voltammograms for oxygen reduction at platinum nanoparticles dispersed within nanocomposites: (A) Au-Ir-rGO, (B) Ir-NPS-rGO, (C) Au-NPS-, (D) rGO. Electrolyte: oxygen-saturated 0.5 moldm$^{-3}$ H$_2$SO$_4$. Scan rate: 10 mVs$^{-1}$. Rotation rate: 1600 rpm. Ring currents are recorded upon application of 1.21V.

**Fig. 11** Percent fraction of hydrogen peroxide (% H$_2$O$_2$) produced during electroreduction of oxygen (and detected at ring at 1.21V) under conditions of the RRDE voltammetric experiments as for Fig. 9 and Fig.10.

**Fig. 12** Numbers of transferred electrons (n) per oxygen molecule during electroreduction of oxygen under under conditions of the RRDE voltammetric experiments as for Fig. 9 and Fig. 10.

Fig. 1

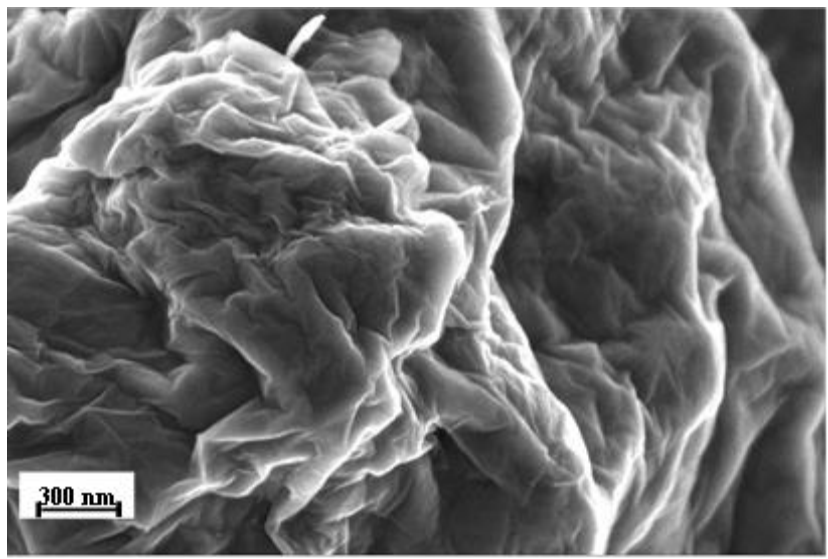

Fig.2

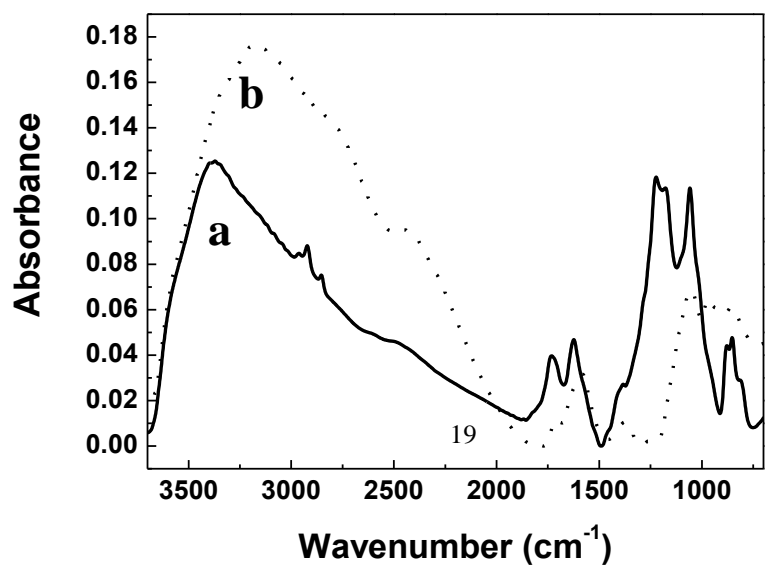

Fig.3

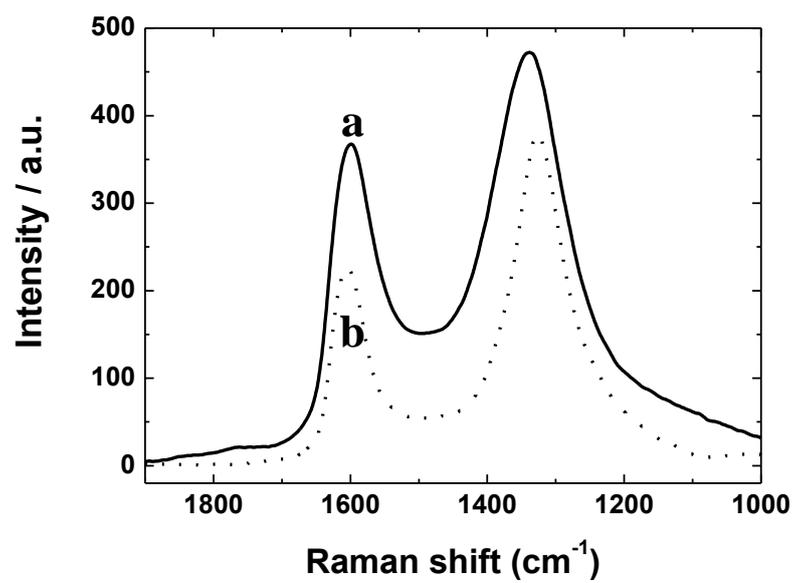

Fig.4

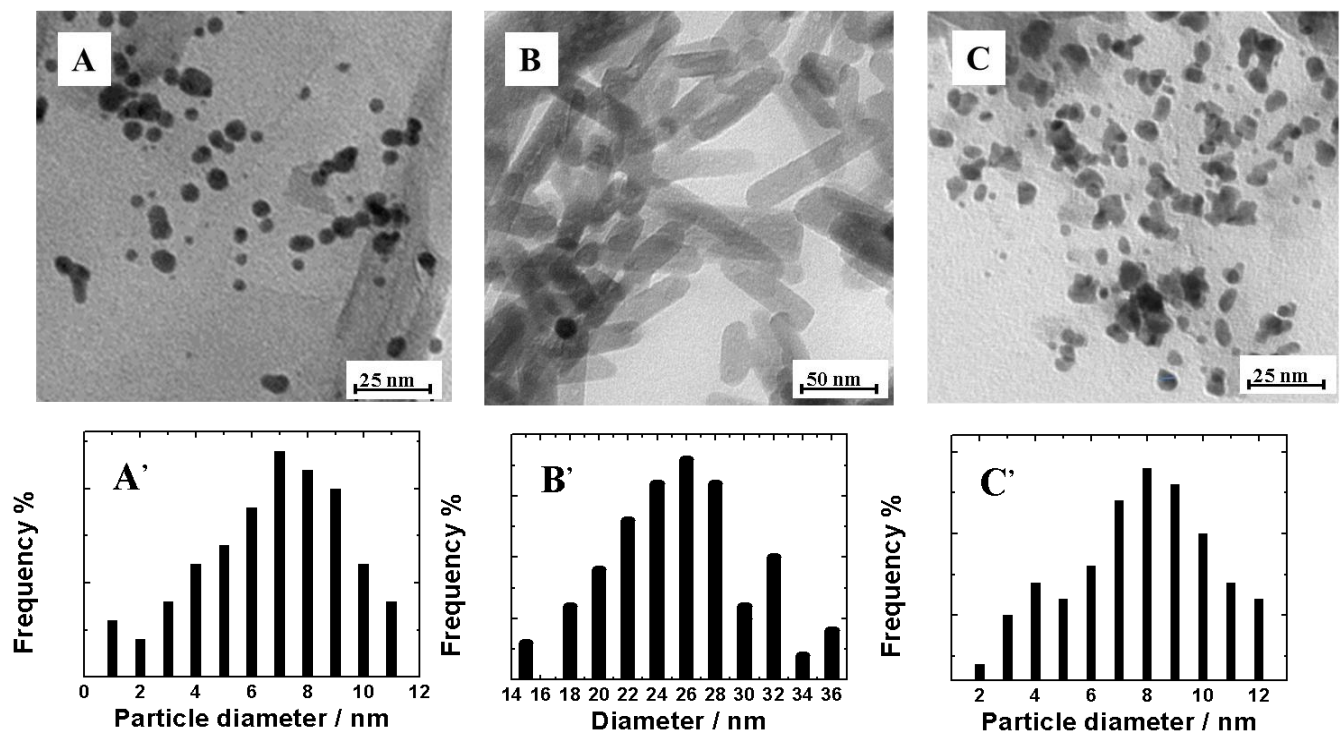

Fig.5

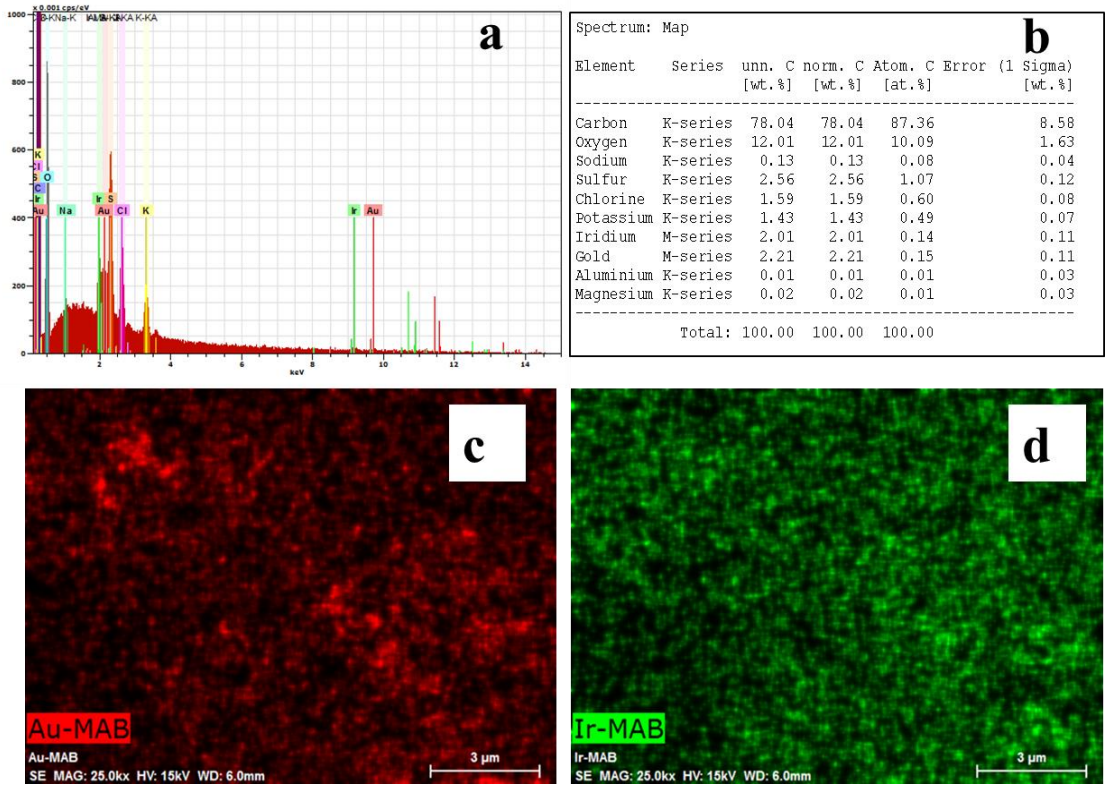

Fig.6

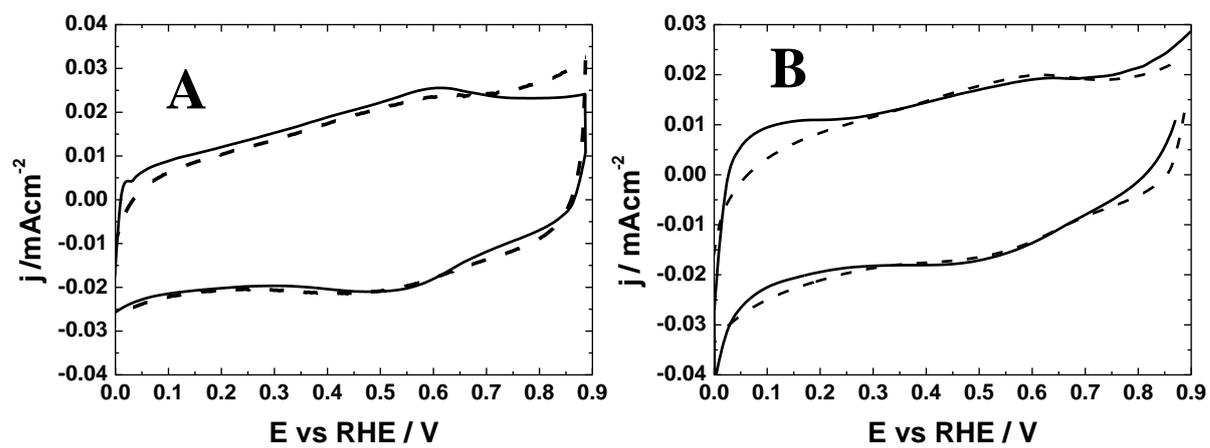

Fig. 7

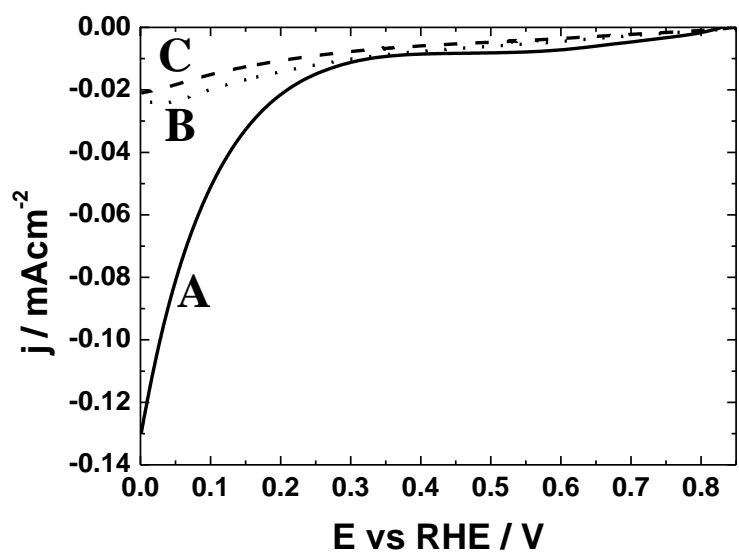

Fig. 8

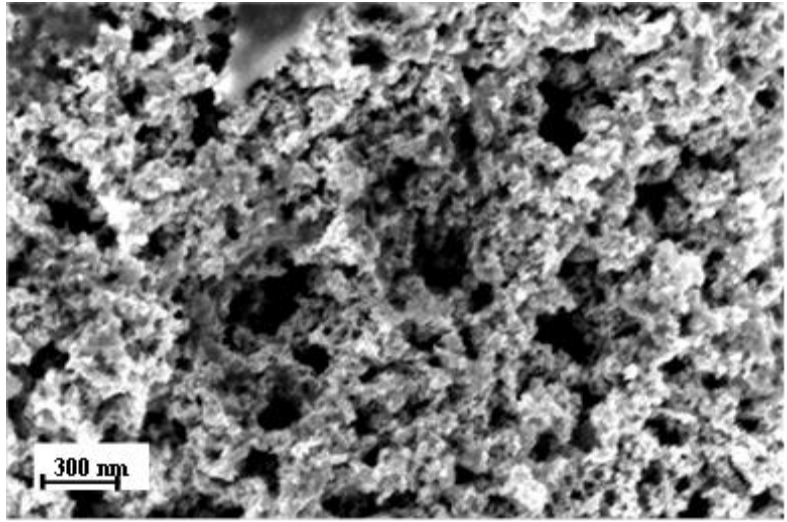

Fig.9

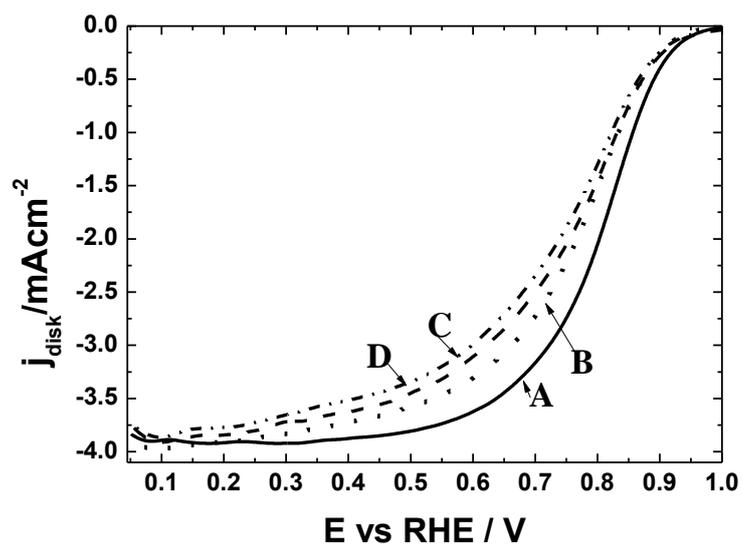

Fig.10

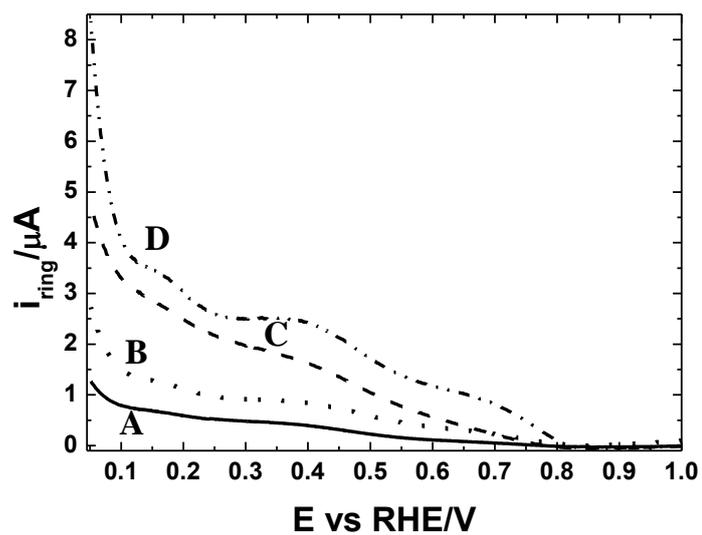

Fig. 11

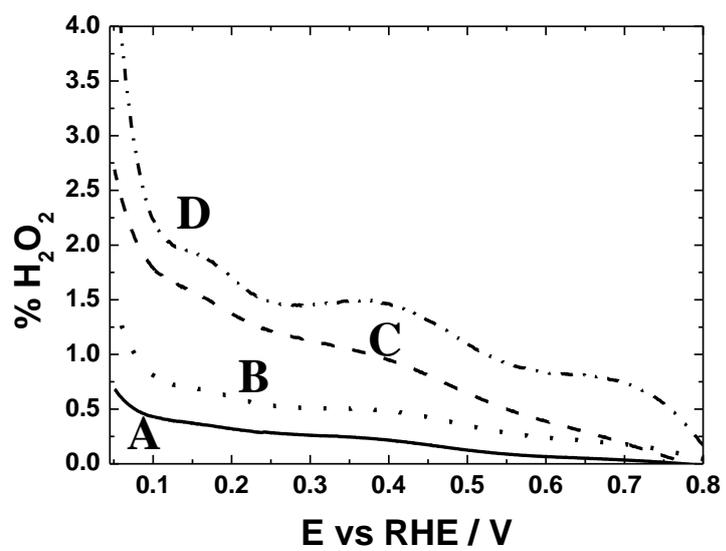

Fig. 12

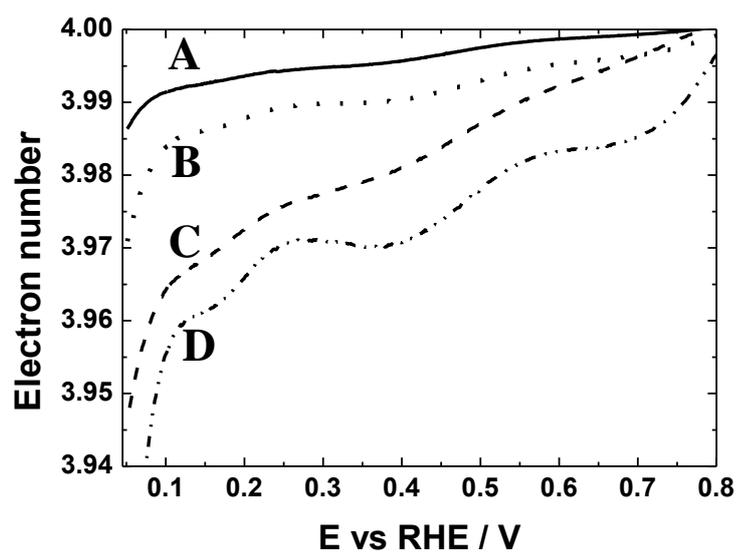